\documentclass[prb,twocolumn,showpacs,preprintnumbers,amsmath,amssymb,floatfix]{revtex4}
\usepackage{color,graphics,epsfig,rotating}
\usepackage{graphicx}
\usepackage{dcolumn}
\usepackage{bm}

\begin{document}

\title{Density functional study of BaNi$_2$As$_2$: Electronic structure,
  phonons and electron-phonon superconductivity}

\author{Alaska Subedi}
\affiliation{Department of Physics and Astronomy, University of Tennessee,
  Knoxville, TN 37996}
\affiliation{Materials Science and Technology Division, Oak Ridge National
  Laboratory, Oak Ridge, Tennessee 37831-6114, USA}

\author{David J. Singh}
\affiliation{Materials Science and Technology Division, Oak Ridge National
  Laboratory, Oak Ridge, Tennessee 37831-6114, USA}

\date{\today}

\begin{abstract}
We investigate the properties of BaNi$_2$As$_2$ using first principles
calculations. The band structure has a similar shape to that of the
BaFe$_2$As$_2$, and in particular shows a pseudogap between a manifold
of six heavy $d$ electron bands and four lighter $d$ bands, i.e. at an
electron count of six $d$ electrons
per Ni. However, unlike BaFe$_2$As$_2$, where the
Fermi energy occurs at the bottom of the pseudogap, the two additional
electrons per Ni in the Ni compound place the Fermi energy in the upper
manifold. Thus BaNi$_2$As$_2$ has large Fermi surfaces very
distinct from BaFe$_2$As$_2$. Results for the phonon spectrum
and electron-phonon
coupling are consistent with a classification of this material as a
conventional phonon mediated superconductor although spin fluctuations
and nearness to magnetism may be anticipated based on the value of
$N(E_F)$.
\end{abstract}


\maketitle

\section{Introduction}

The finding of superconductivity in F doped LaFeAsO by Kamihara and co-workers
\cite{kamihara} has led to considerable interest as this provides
an alternative non-cuprate route to high critical temperature ($T_c$)
superconductivity.
Since this finding,
high $T_c$ superconductivity has been found in
many related phases that can be generally characterized into four different
groups according to their parent compounds: (i)
LaFeAsO,\cite{kamihara,chen,zhu} (ii) BaFe$_2$As$_2$,\cite{rotter1,rotter2}
(iii) LiFeAs,\cite{wang:Li} and (iv) $\alpha$-FeSe.\cite{hsu} The
precise mechanism of
superconductivity in these compounds is yet to be
established, but is strongly thought to be
unconventional.
This is based in part on calculations
of the electron-phonon coupling, \cite{boeri,mazin}
which is too small to account for
any appreciable superconductivity and on
the proximity to magnetism. \cite{cruz,singh3,huang,mazin2,chen2}
These compounds show considerable variation in
doping levels (i.e. hole or electron doped),
applied pressure, and chemistry of the non-Fe layers,
while remaining superconducting. However, there are three main threads
joining these compounds together. The parent compounds have spin density wave
(SDW) order with the possible
exception of the LiFeAs phase,
exist in a tetragonal structure and possess iron in a 2D square lattice
sheet. There is a clear association between suppression of the SDW order
and appearance of superconductivity in the phase diagrams, although
recent work does show some range of
coexistence of magnetism and superconductivity.
\cite{chenh,drew}
Importantly, the SDW phase is accompanied by an orthorhombic distortion.
Depending on the specific material, this distortion occurs at a temperature
higher than the SDW ordering temperature or coincident with it.
\cite{cruz,rotter1,yan}

In addition, Ronning \textit{et al.}\ recently reported
both the occurrence of a first order phase
transition at $T_0 = 130$ K with characteristics similar to the
structural transition seen in the Fe-As based superconductors,
and a superconducting transition with $T_c = 0.7$ K in
BaNi$_2$As$_2$.\cite{ronning}
Since Fe and Ni are ambient temperature
ferromagnets and many Fe and Ni compounds show magnetism, it is plausible to
expect that Ni can fill in the role of Fe in these compounds. Indeed,
superconductivity is also observed in LaNiPO,\cite{watanabe} LaNiAsO,\cite{li}
and LaNiBiO.\cite{kozhevnikov} However, these compounds do not display the
structural or magnetic transitions characteristic of the Fe-As based parent
compounds.
Moreover, it was recently shown that LaNiPO can be
explained as a conventional electron-phonon superconductor.\cite{subedi}

In this paper we report the details of our calculations of the electronic
structure, phonons and electron-phonon coupling of BaNi$_2$As$_2$. The band
structure of BaNi$_2$As$_2$ is similar to that reported
for BaFe$_2$As$_2$, \cite{nekrasov-basr,ma-ba,singh}
but the Fermi level is shifted up owing to the higher
valence electron count in Ni$^{2_+}$ than in Fe$^{2+}$. Hence, with a larger
Fermi surface and higher carrier density, BaNi$_2$As$_2$ has remarkably
different electronic properties from BaFe$_2$As$_2$. A similar increase
in the Fermi surface area is also found in LaNiPO for
essentially the same reason.\cite{subedi} Also similar to
case of LaNiPO, we obtain a much
larger value of electron-phonon coupling constant
$\lambda_{ep} = 0.76$ in BaNi$_2$As$_2$ compared to the Fe based compounds (for
example, in LaFeAsO $\lambda_{ep} = 0.21$\cite{boeri}). This suggests that the
mechanism of superconductivity in BaNi$_2$As$_2$ is similar to that of LaNiPO
and is different from the Fe based compounds.

\section{Methods and Structure}

The electronic structure calculations were performed within the local density
approximation (LDA) with the general potential linearized augmented planewave
(LAPW) method,\cite{singh2} similar to the calculations reported for
LaFeAsO.\cite{singh3} We used LAPW spheres of radius 2.2 $a_0$ for Ba and 2.1
$a_0$ for Ni and As. BaNi$_2$As$_2$ occurs in a body centered tetragonal
structure ($I4/mmm$) with Ba, Ni and As at the positions 2a(0,0,0),
4d(0.5,0,0.5) and 4e(0,0,$z_{\rm As}$), respectively. Here $z_{\rm As}$, the
As height above the Ni square planes, is a structural parameter governing
the Ni-As distance and the distortion of the As tetrahedra that
coordinate the Ni in this material.
In our calculations, we used experimental lattice
parameters ($a = 4.112$ \AA\ and $c = 11.54$ \AA)\cite{ronning} but employed
the computed $z_{\rm As}$ obtained via non-spin polarized energy minimization.

The phonon dispersions and electron-phonon coupling were calculated using
linear response as implemented in Quantum Espresso code,\cite{espresso}
similar to the calculations reported for LaFeAsO and LaNiPO.
\cite{boeri,mazin,subedi}
The linear response calculations were also done with
experimental lattice parameters, using ultrasoft pseudopotentials within the
generalized gradient approximation (GGA)
of Perdew, Burke and Ernzerhof.\cite{pbe}
An 8x8x8 grid was used for the zone integration in the phonon calculations,
while a more dense 32x32x8 grid was used for the zone integration in the
electron-phonon coupling calculations. The basis set cut-off for the wave
functions was 40 Ry, while a 400 Ry cut-off was used for the charge density.

The internal parameter $z_{\rm As}$ was again relaxed in the calculation of
phonon properties. The values we obtained for $z_{\rm As}$ (LDA: 0.346, GGA:
0.351) agree well with the experimental value of $z_{\rm As} = 0.3476$. This
is in contrast to BaFe$_2$As$_2,$\cite{singh} and other Fe based
superconductors where LDA and GGA
calculations done in this way noticeably underestimate $z_{\rm As}$,
but is similar to the case of LaNiPO. It has been suggested that this
underestimation of $z_{\rm As}$ in Fe based superconductors is due to its
coupling with magnetism.\cite{mazin2,yin,mazin3}
This underestimation may be indicative of strong spin fluctuations
in the paramagnetic superconducting parts of the phase diagrams
of the Fe-based materials.
The absence of this discrepancy between the
experimental and calculated value of $z_{\rm As}$ in BaNi$_2$As$_2$
then indicates
that the magnetic character of BaNi$_2$As$_2$ is different from that of
BaFe$_2$As$_2$ and other Fe based superconductors.

\begin{figure}[tbp]
  \includegraphics[scale=0.3,angle=270]{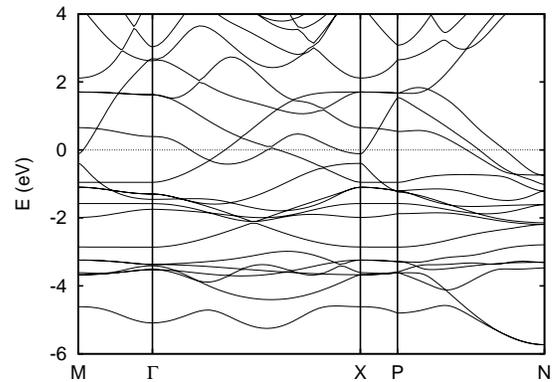}
  \caption{Calculated LDA band structure of non-spin-polarized
    BaNi$_2$As$_2$.}
  \label{fig:BaNi2As2-bnd}
\end{figure}

\begin{figure}[tbp]
  \includegraphics[scale=0.35,angle=270]{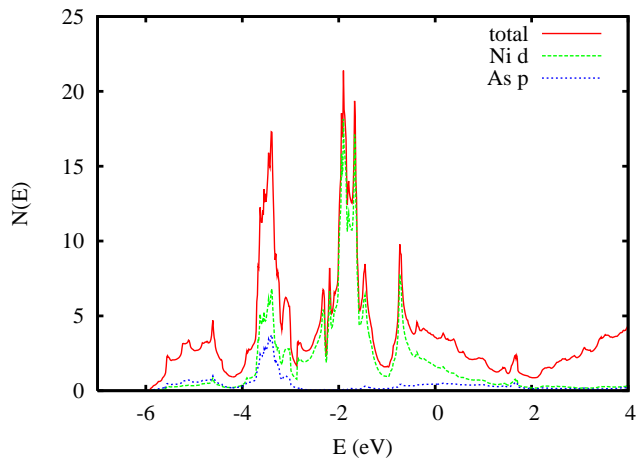}
  \caption{(color online)
Calculated LDA electron density of states of non-spin-polarized
    BaNi$_2$As$_2$ and projections onto the LAPW spheres on a per
formula unit basis.}
  \label{fig:BaNi2As2-dos}
\end{figure}

\begin{figure}[tbp]
  \includegraphics[width=0.9\columnwidth]{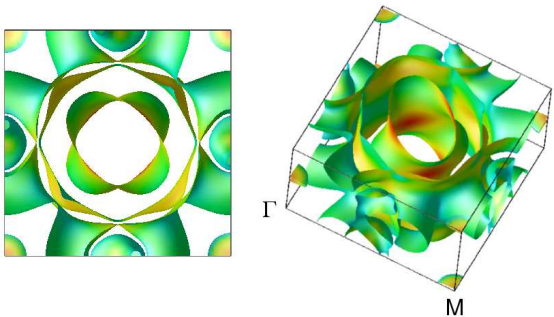}
  \caption{(color online) Calculated LDA Fermi surface of non-spin-polarized
    BaNi$_2$As$_2$. The shading is by velocity.}
  \label{fig:BaNi2As2-fs}
\end{figure}

\begin{figure}[tbp]
  \includegraphics[scale=0.4]{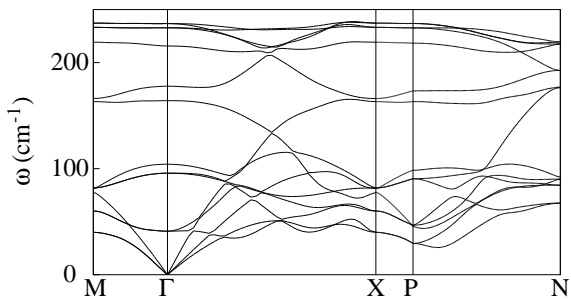}
  \caption{Calculated GGA phonon dispersion curves of non-spin-polarized
    BaNi$_2$As$_2$.}
  \label{fig:BaNi2As2-phon}
\end{figure}

\begin{figure}[tbp]
  \includegraphics[scale=0.3,angle=270]{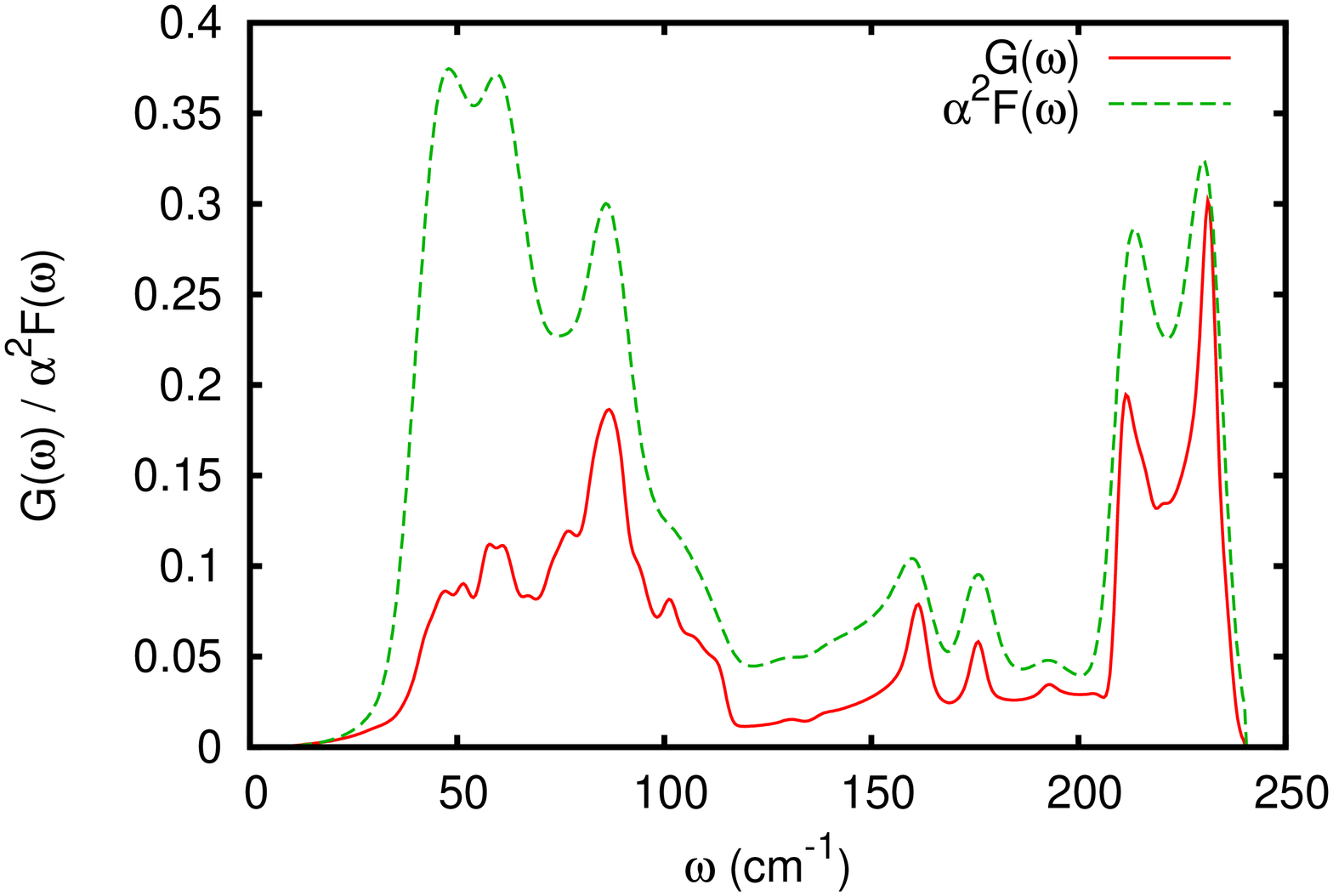}  
  \includegraphics[scale=0.3,angle=270]{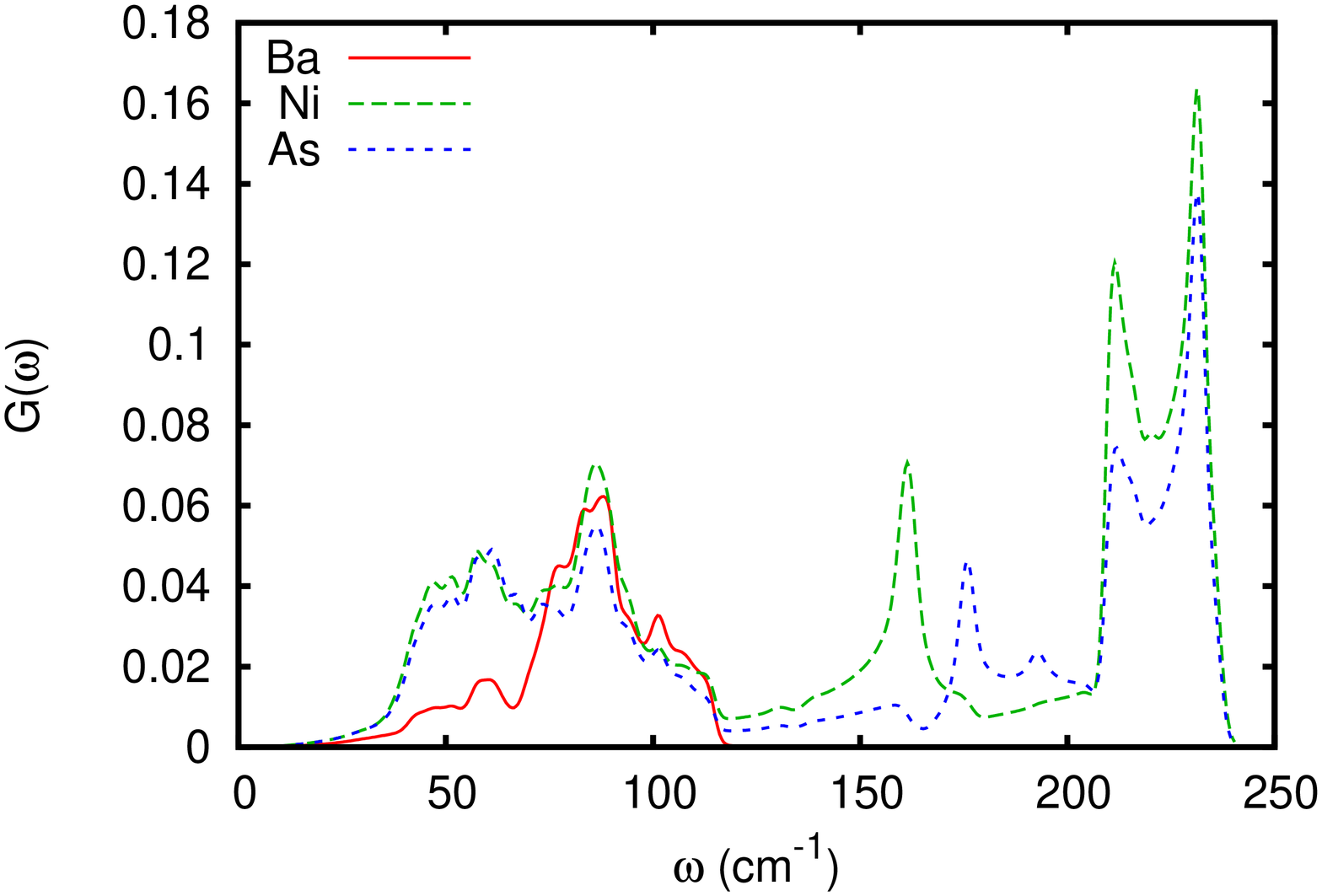}
  \includegraphics[scale=0.3,angle=270]{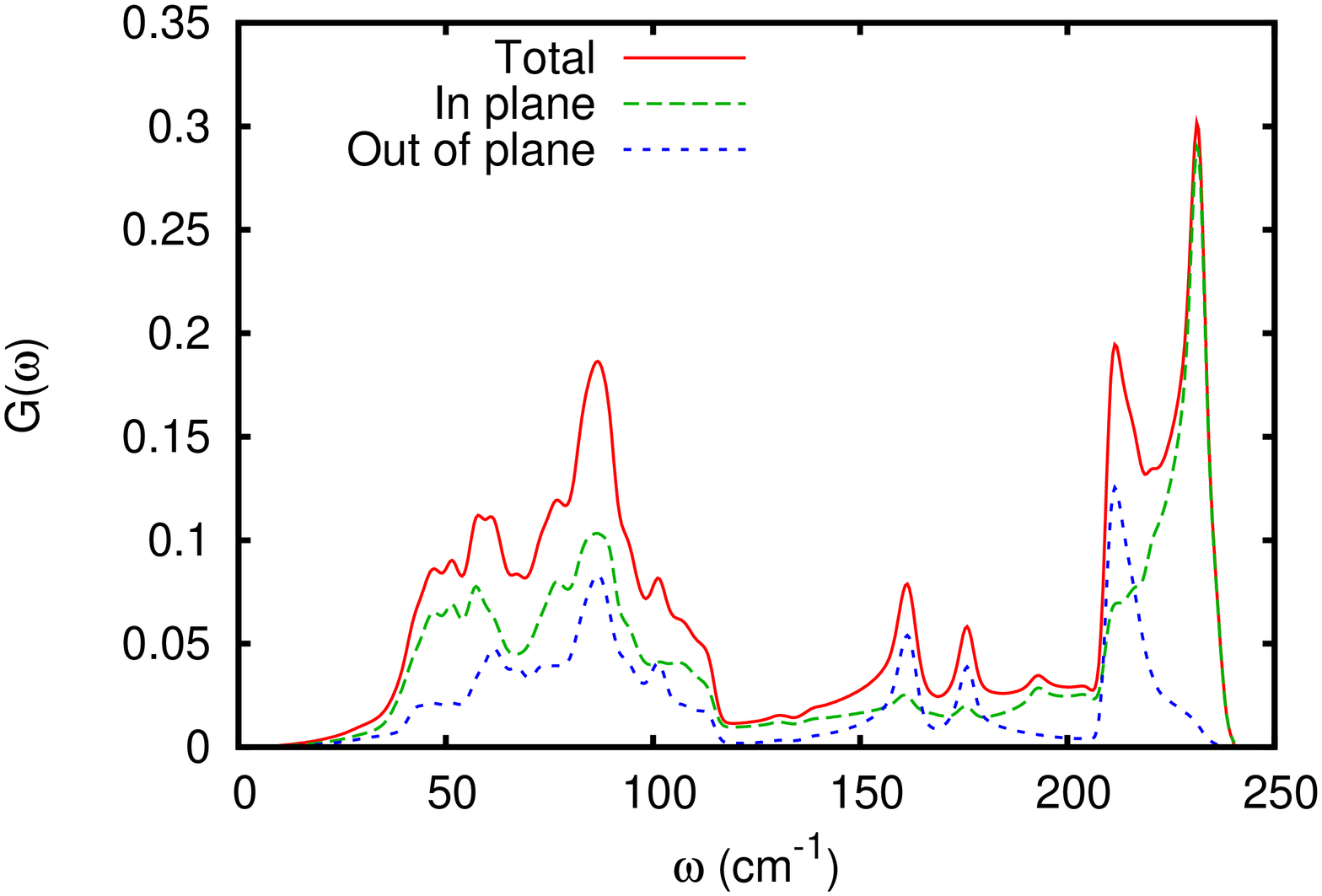}
  \caption{(color online)
Top: Calculated GGA phonon density of states $G(\omega)$ and
    electron-phonon spectral function $\alpha^2F(\omega)$ for
    non-spin-polarized BaNi$_2$As$_2$. Middle: Projected phonon density of
    states for each atom of BaNi$_2$As$_2$. Bottom: Phonon density
    of states weighted by in-plane or out-of-plane character, i.e.
atomwise projections of eigenvectors in the $ab$-plane and along
the $c$-axis, respectively.}
  \label{fig:pdos-a2F}
\end{figure}

\section{Results}

The calculated band structure and electronic density of states (DOS) are shown
in Figs. \ref{fig:BaNi2As2-bnd} and \ref{fig:BaNi2As2-dos}, respectively. The
Fermi surface is shown in Fig. \ref{fig:BaNi2As2-fs}.
The pnictogen $p$ states occur between -6 eV and -3 eV, with respect
to the Fermi energy, consistent with an anionic As$^{3-}$ species, and
modest hybridization of the
As $p$ states with Ni $d$ states similar to the Fe-based
materials.
As expected, Ba occurs as Ba$^{2+}$ with the Ba derived valence states
well above the Fermi energy. As such, Ni has a nominal valence Ni$^{+2}$
with eight $d$ electrons per Ni.
The manifold of six heavy bands between
-3.0 eV and -1.0 eV have Ni $d$ character and account for six electrons per Ni.
There is a pseudogap at this point separating the heavy bands from a manifold
of
lighter bands that span between -1.0 eV and 2.0 eV.
The light bands accommodate the remaining $d$ electrons
and show Ni $d$ character accompanied by some mixing with As $p$
states. It may seen that the band structure and DOS of BaNi$_2$As$_2$
are qualitatively very similar to that of BaFe$_2$As$_2$ 
(Refs. \onlinecite{nekrasov-basr},\onlinecite{ma-ba}, and \onlinecite{singh}),
with the
exception that Ni$^{2+}$ crucially contains two more
valence electrons than Fe$^{2+}$. This causes the Fermi level to shift up away
from the pseudogap into the upper manifold. As a result, compared to
BaFe$_2$As$_2$, BaNi$_2$As$_2$ has a large
multi-sheet Fermi surface very different from
the Fe-based materials.
The value of the DOS at $E_F$ is $N(E_F)$=3.57 eV$^{-1}$,
on a per formula unit (two Ni atoms) both spins basis.
This is lower than LaFeAsO, but is comparable to some of the other
Fe-As based materials.

The calculated phonon dispersions of BaNi$_2$As$_2$ are shown in
Fig.~\ref{fig:BaNi2As2-phon}. The corresponding phonon density of states and
Eliashberg spectral function $\alpha^2F(\omega)$ are shown in
Fig.~\ref{fig:pdos-a2F}. There are 15 phonon bands extending up to $\sim$ 230
cm$^{-1}$. The bands below 120 cm$^{-1}$ show Ba, Ni and As characters while
the region above it contains bands of mostly Ni and As character.
From Fig.~\ref{fig:pdos-a2F}, we can see that the
electron-phonon spectral function is
enhanced relative to the phonon density of states in the low frequency
modes. The projected phonon density of states plots (Fig.~\ref{fig:pdos-a2F},
middle and bottom) show that this enhancement occurs in the region where there
is high Ni and As character, although the enhancement cannot be attributed to
solely in-plane or out-of-plane character. This is in contrast to the case of
LaFeAsO where the spectral function more closely follows
in proportionality to the
phonon density of states in this energy region.\cite{boeri}
It should be noted that doped LaFeAsO was shown to have
a rather low overall electron-phonon
coupling ($\lambda_{ep} \sim 0.2$) which cannot explain the
superconductivity\cite{boeri} while LaNiPO was shown to be be
readily explained as conventional electron-phonon superconductor.\cite{subedi}
For
BaNi$_2$As$_2$,
we obtain a value of the electron-phonon coupling $\lambda_{ep} =
0.76$ with logarithmically average frequency $\omega_{\ln}=73$
cm$^{-1}$. Inserting these numbers into the simplified Allen-Dynes formula,
\begin{equation}
k_B T_c = {\frac{\hbar \omega_{\rm ln}}{1.2}} ~~ {\rm exp}
\left \{
- {\frac{1.04 (1 + \lambda_{ep})} {\lambda_{ep} - \mu^* (1 + 0.62 \lambda_{ep})}}
\right \} ,
\end{equation}
with $\mu^*=0.12$, we obtain $T_c \sim$ 4K, which overestimates but is
reasonably in accord with the experimental low value of $T_c =$
0.7K.\cite{ronning}
As mentioned, the total DOS per formula unit at the Fermi energy is
$N(E_F)$ = 3.57 eV$^{-1}$.
With a typical $3d$
Stoner parameter $I \sim$ 0.7 -- 0.8 eV, this yields
a Stoner enhancement $(1-NI)^{-1} \sim$  3
(note that in this formula $N$ is per atom per spin).
This enhancement is sufficient to indicate the presence of spin fluctuations
that would depress the electron-phonon $T_c$ and may therefore explain
why the experimental
$T_c$ is reduced compared with the calculated value based on $\lambda_{ep}$.

\section{Conclusions}

In summary, we have presented electronic structure calculations that show
BaNi$_2$As$_2$ has very different electronic properties compared to the
Fe based
high $T_c$ superconductors. Even though it has a band structure
similar to that of BaFe$_2$As$_2$, the Fermi level lies in the upper manifold
away from the pseudogap due to higher number of valence electrons in
Ni$^{2+}$. This gives BaNi$_2$As$_2$ a large Fermi surface in
contrast to the small surfaces in the Fe
based superconductors. 
We obtain a moderately high $N(E_F)$= 3.57 eV$^{-1}$,
which would yield a Stoner renormalization of $\sim$ 3, consistent with
spin fluctuations that would
reduce the electron-phonon $T_c$.
Nonetheless,
the calculated value for electron-phonon coupling constant is
$\lambda_{ep} = 0.76$, which is ample for a description of BaNi$_2$As$_2$ as an
electron-phonon superconductor similar to LaNiPO.

\acknowledgements

We are grateful for helpful discussions with M.H. Du and I.I. Mazin.
This work was supported by the Department of Energy,
Division of Materials Sciences and Engineering.


\begin{references}

\bibitem{kamihara}
  Y. Kamihara, T. Watanabe, M. Hirano, and H. Hosono, 
  J. Am. Chem. Soc. {\bf 130}, 3296 (2008).

\bibitem{chen} 
  G. F. Chen, Z. Li, G. Li, J. Zhou, D. Wu, J. Dong, W. Z. Hu,
  P. Zheng, Z. J. Chen, H. Q. Yuan, J. Singleton, J. L. Luo, N. L. Wang,
  Phys. Rev. Lett. {\bf 101}, 057007 (2008).

\bibitem{zhu}
  X. Zhu, H. Yang, L. Fang, G. Mu, H.H. Wen,
  Supercond. Sci. Technol. {\bf 21} 105001 (2008).

\bibitem{rotter1}
  M. Rotter, M. Tegel, D. Johrendt, I. Schellenberg, W. Hermes,
and R. Pottgen,
  Phys. Rev. B {\bf 78}, 020503(R) (2008).

\bibitem{rotter2}
  M. Rotter, M. Tegel, and D. Johrendt,
  arXiv:0805.4630v2 (2008).

\bibitem{wang:Li} 
  X.C.Wang, Q.Q. Liu, Y.X. Lv, W.B. Gao, L.X.Yang, R.C.Yu, F.Y.Li, and C.Q. Jin,
  arXiv:0806.4688v2 (2008).

\bibitem{hsu} 
  F.C. Hsu, J.Y. Luo, K.W. Yeh, T.K. Chen, T.W.  Huang, P. M. Wu,
  Y.C. Lee, Y.L. Huang, Y.Y Chu, D.C. Yan, and M.K. Wu,
  arXiv:0807.2369v2 (2008).

\bibitem{boeri}
  L. Boeri, O.V. Dolgov, and A.A. Golubov,
  Phys. Rev. Lett. {\bf 101}, 026403 (2008).

\bibitem{mazin}
I.I. Mazin, D.J. Singh, M.D. Johannes and M.H. Du,
Phys. Rev. Lett. {\bf 101}, 057003 (2008).

\bibitem{cruz}
C. de la Cruz, Q. Huang, J.W. Lynn, J. Li, W. Ratcliff II, J.L. Zaretsky,
H.A. Mook, G.F. Chen, J.L. Luo, N.L. Wang, and P. Dai,
Nature (London) {\bf 453}, 899 (2008).

\bibitem{singh3}
  D.J. Singh and M.H. Du, 
  Phys. Rev. Lett. {\bf 100}, 237003 (2008). 

\bibitem{huang}
Q. Huang, J. Zhao, J.W. Lynn, G.F. Chen, J.L. Luo, N.L. Wang, and P. Dai,
Phys. Rev. B {\bf 78}, 054529 (2008).

\bibitem{mazin2}
I.I. Mazin and M.D. Johannes, arXiv:0807.3737 (2008).

\bibitem{chen2}
G.F. Chen, Z. Li, D. Wu, G. Li, W.Z. Hu, J. Dong, P. Zheng, J.L. Luo,
and N.L. Wang,
Phys. Rev. Lett. {\bf 100}, 247002 (2008).

\bibitem{chenh}
H. Chen, Y. Ren, Y. Qiu, W. Bao, R.H. Liu, G. Wu, T. Wu, Y.L. Xie, X.F. Wang,
Q. Huang, and X.H. Chen, 
arXiv:0807.3950 (2008).

\bibitem{drew}
A.J. Drew, C. Niedermayer, P.J. Baker, F.L. Pratt, S.J. Blundell,
R.H. Lancaster, R.H. Liu, G. Wu, X.H. Chen, I. Watanabe, V.K. Malik,
A. Dubroka, M. Roessle, K.W. Kim, C. Baines, and C. Bernhard,
arXiv:0807.4876 (2008).

\bibitem{yan}
J.Q. Yan, A. Kreyssig, S. Nandi, N. Ni, S.L. Budko, A. Kracher, R.J. McQueeney,
R.W. McCallum, T.A. Lograsso, A.I. Goldman, and P.C. Canfield,
Phys. Rev. B {\bf 78}, 024516 (2008).



\bibitem{ronning}
  F. Ronning, N. Kurita, E.D. Bauer, B.L. Scott, T. Park, T. Klimczuk,
  R. Movshovich, and J.D. Thompson,
  arXiv:0807.3788, (2008). 

\bibitem{watanabe} 
  T. Watanabe, H. Yanagi, T. Kamiya, Y. Kamihara, H. Hiramatsu, M.  Hirano,
  and H. Hosono, 
  Inorg. Chem. {\bf 46}, 7719 (2007).
  
\bibitem{li} 
  Z. Li, G. F. Chen, J. Dong, G. Li, W. Z. Hu, D. Wu, S. K. Su, P. Zheng,
  T. Xiang, N. L. Wang, J. L. Luo,
  Phys. Rev. B, {\bf 78}, 060504(R) (2008).

\bibitem{kozhevnikov}
  V.L. Kozhevnikov, O.N. Leonidova, A.L. Ivanovskii, I.R Shein,
  B.N. Goshchitskii, and A.E. Karkin,
  arXiv:0804.4546v1 (2008).

\bibitem{subedi}
  A. Subedi, D.J. Singh, and M.H. Du,
  Phys. Rev. B {\bf 78}, 060506(R) (2008).

\bibitem{nekrasov-basr}
I.A. Nekrasov, Z.V. Pchelkina, and M.V. Sadovskii,
arXiv:0806.2630 (2008).

\bibitem{ma-ba}
F. Ma, Z.Y. Lu, and T. Xiang,
arXiv:0806.3526 (2008).

\bibitem{singh}
  D.J. Singh
  arXiv:0807.2643 (2008) 

\bibitem{singh2}
  D.J. Singh and L. Nordstrom,
  {\it Planewaves, Pseudopotentials and the LAPW Method, 2nd.\ Ed.} (Springer,
  Berlin, 2006).

\bibitem{espresso}
  S. Baroni, A. Dal Corso, S. de Gironcoli, P. Gianozzi, C. Cavazzoni,
  G. Ballabio, S. Scandolo, G. Chiarotti, P. Focher, A. Pasquarello, {\it et
    al.},
  http://www.quantum-espresso.org.

\bibitem{pbe}
  J.P. Perdew, K. Burke, and M. Ernzerhof,
  Phys. Rev. Lett. {\bf 77}, 3865 (1996). 
  
\bibitem{yin}
  Z. P. Yin, S. Leb\`egue, M. J. Han, B.P. Neal, S. Y. Savrasov, W. E. Pickett,
  Phys. Rev. Lett. {\bf 101}, 047001 (2008)

\bibitem{mazin3}
I.I. Mazin, M.D. Johannes, L. Boeri, K. Koepernik, and D.J. Singh,
Phys. Rev. B {\bf 78}, 085104 (2008).

\end{references}
\end{document}